**Field-material coupled neural network: A novel prior-free and data-free inverse problem solver for extracting complex dielectric constant in terahertz band**

Pengfei Zhu[1,*], Stefano Sfarra[2], Elena Pivarciova[3], Carlo Santulli[4], Xavier Maldague[5]

[1] *Bundesanstalt für Materialforschung und -prüfung (BAM), 12200, Berlin, Germany*
[2] *Department of Industrial and Information Engineering and Economics (DIIIE), University of L'Aquila, I-67100 L'Aquila, Italy*
[3] *Faculty of Environmental and Technology Manufacturing, Technical University in Zvolen, 960 53, Zvolen, Slovakia*
[4] *Geology Division, School of Science and Technology (SST), Università degli Studi di Camerino, Camerino, Italy*
[5] *Department of Electrical and Computer Engineering, Computer Vision and Systems Laboratory (CVSL), Laval University, Quebec G1V 0A6, Canada*

[*]Corresponding author: pengfei.zhu@bam.de

**ABSTRACT**

Accurate extraction of the complex dielectric constant in the terahertz (THz) band is essential for material characterization and non-destructive evaluation yet remains challenging due to the ill-posed nature of electromagnetic inverse problems and the limited availability of reliable reference data. In this work, a field-material couple neural network (FMCNN) is proposed to retrieve the complex dielectric constant directly from THz measurements. The FMCNN consists of a field neural network and a material neural network that are strongly coupled through the frequency-domain Maxwell equations in the form of a Helmholtz equation, with the governing physics enforced by partial differential equation (PDE) and boundary condition constraints. This formulation enables prior-free and data-free inversion, requiring only measured test data as input. The extracted dielectric constants are validated by comparison with results from a one-dimensional normal-incidence model and the Drude-Lorentz model, showing good agreement over a broad frequency range, particularly above 0.2 THz. These results demonstrate that the FMCNN provides a physics-consistent and data-efficient approach for material parameter extraction in the THz band, offering an alternative to conventional model-based methods.

Keywords: THz-TDS, terahertz, refractive index, permittivity, physics-informed neural network (PINN)

**I. INTRODUCTION**

The terahertz band of the electromagnetic spectrum spans from ~100 GHz (3 mm wavelength) to ~10 THz (30 μm). Historically, this band was hard to access[1]. Unlike lower frequencies, where radiation is efficiently generated and detected using electronic circuits, or higher frequencies, which rely on quantum transitions in atoms, molecules, or solids, the THz regime lies between these two well-established domains, limiting the applicability of conventional electronic and optical technologies[2-4]. Extracting the optical and electrical parameters of materials in the terahertz frequency range is of fundamental and practical importance, as these parameters govern carrier dynamics, polarization response, and low-energy excitations that are inaccessible at optical or microwave frequencies[5]. However, there are currently few efficient, non-destructive methods for obtaining the optical and electrical properties in THz band[6].

The most common method before terahertz time-domain spectroscopy (THz-TDS) for broadband access to the terahertz spectral range is using a thermal blackbody as the radiation source. However, these sources do not produce much radiation at frequencies below a few terahertz. Then, the first and most widespread achieving THz generation and detection with extraordinary signal-to-noise ratio (SNR) is THz-TDS systems[7,8]. In this case, the THz generation includes photoconductive emitters[9] and optical rectification[10]. A photoconductive emitter consists of a pair of metallic electrodes with a gap on a highly resistive semiconductor substrate such as low temperature-grown GaAs[11]. A large DC electric field is applied across the gap between two electrodes. When an ultrashort laser pulse goes through the gap between the electrodes, the generated photo-carriers will accelerate in the applied DC field. The rapidly varying current $J$ gives rise to a burst of electromagnetic (EM) radiation $E_{\text{THz}}$, in accordance with Maxwell's equations $E_{\text{THz}} \propto \frac{\partial J}{\partial t} = \frac{\partial}{\partial t}(nev)$. Where $n$ is the photo-carrier density, $e$ is the electron charge, and $v$ is the charge velocity. Optical rectification requires a material with non-zero bulk second-order dielectric susceptibility[12]. When a monochromatic wave with a Gaussian temporal profile and a frequency $\omega$ travels through the material, a second-order dielectric polarization is generated according to $P^{(2)} = \frac{1}{2}\chi^{(2)}E_0^2(1-\cos(2\omega t))e^{-(t/\tau)^2}$ (where $\chi^{(2)}$ is the second-order susceptibility of the material, $\tau$ is proportional to the duration of the pulse). The first term (zero-frequency polarization term, $P_0 = \frac{1}{2}\chi^{(2)}E_0^2 e^{-(t/\tau)^2}$) produces an EM pulse $E_{\text{THz}} \propto \frac{d^2 P}{dt^2}$. The optical rectification technique provides much larger bandwidth but lower emitted THz power comparing with photoconductive emitters. THz detection includes photoconductive receivers[13] and electro-optic sampling[14]. The photoconductive antenna in photoconductive receivers is same as that in photoconductive emitters. The difference is to replace the bias voltage with a photocurrent. An ultrashort laser pulse synchronized with THz transients is focused on the gap between the two contacts. The laser pulse is absorbed, rapidly changing the conductivity of the semiconductor due to the generation of photo-carriers. Then, a measurable current is generated by the electric field of the incoming THz pulse according to $I(t) = \frac{\mu e P_\text{G} T_{12} \lambda}{hcW} \int_{-\infty}^{+\infty} E_{\text{THz}}(t)\emptyset(t-\tau)dt$[15]. Where $\tau$ is the relative delay between the THz and the optical pulse, $\mu$ is the mobility in the semiconductor, $P_\text{G}$ is the optical power incident on the receiver, $T_{12}$ is the Fresnel transmission coefficient, $\lambda$ is the central wavelength of the laser, $h$ is Planck's constant, $c$ is the speed of light, $W$ is the laser beam width, and $\emptyset(t-\tau)$ is the temporal response (gating) function of the photoconductive receiver. Another THz detection modality is electro-optic sampling, which is based on the Pockels effect. A THz electric field induces an ultrafast birefringence in a nonlinear crystal. An ultrashort optical probe pulse senses this birefringence via its polarization change, enabling time-resolved sampling of the THz electric field[16]. In general, electro-optic sampling offers higher bandwidth and more linear, distortion-free detection of THz fields, while PCA detection provides higher sensitivity and simpler implementation but is limited by carrier lifetime and bandwidth.

THz-TDS can directly and non-invasively probe optical and electronic properties such as the refractive index $n$, absorption coefficient $\alpha$, dielectric constant $\varepsilon$, and conductivity $\sigma$. This has been employed in materials like semiconductors[17], woods[18], fiber reinforced composites[19], low-dimensional materials[20], and metamaterials[21]. Obtaining material parameters requires multiple analysis stages including phase unwrapping, jitter correction, and transfer function convergence, with each step introducing potential errors[22,23]. In 2016, a study compared the refractive index and the absorption coefficient determined by THz-TDS for a set of standard materials[22]. The measurements were performed by different laboratories worldwide, using different THz-TDS systems. The study revealed variation between results, highlighting the need for a standardized measurement technique[22,24]. Much work has been carried out to improve the robustness of THz-TDS extraction. For instance, researchers proposed a timing jitter correction algorithm[25]. However, new errors were introduced, and more computationally demanding extraction process was required.

Nowadays, deep learning has been applied to different scientific fields. In THz-TDS, deep learning is commonly used in feature extraction[26], object classification[27], defect recognition[28], image denoising[29], and super-resolution imaging[30]. Recently, Ben et al.[31] applied artificial neural networks to extract the complex conductivity of two-dimensional materials. However, the supervised learning strategy requires sufficient training datasets with accurate labels. It is not yet general practice to make raw or processed THz-TDS data publicly available. Simulation data can be an alternative. However, conventional simulation modalities including finite-element method[32] and finite-difference time-domain method[33] are generally forward modelling approaches. The frequency-dependent material properties are required for calculation. This renders training strategies based on simulation datasets self-contradictory. Physics-informed neural network (PINN)[34] is a new deep learning technique which embeds physical loss into the neural network for training. It has been applied for the simulation of heat transfer[35], fluid mechanics[36], and solid mechanics[37]. In addition, one of advantages in PINN is solving inverse problems, such as the extraction of material parameters[38]. This motivates us to apply physics-informed learning to extract material intrinsic properties as it does not require a labelled training dataset. However, different from heat transfer, solid mechanics, and fluid mechanics, the material parameter in THz band is frequency-dependent, i.e., the material parameter is a vector / matrix instead of a constant. In addition, the temporal scale in THz band is extremely small, which significantly affects the training of the PINN.

In this work, a field-material coupled neural network (FMCNN) is proposed as a prior-free and data-free inverse solver for extracting the complex dielectric constant in the THz band. Unlike conventional supervised learning approaches for material parameter extraction, the FMCNN treats the neural network solely as a solver, using measured data comprising a reference signal and a transmitted signal as input. During training, the network weights are iteratively optimized to minimize the combined loss. The extracted results are validated by comparison with conventional one-dimensional normal-incidence models and the Drude-Lorentz model.

## II. MATERIAL PARAMETER EXTRACTION

The material parameter extraction via THz-TDS systems depends on the absorption ability and thickness of the tested sample. In general, transmission mode is used to extract the material parameter as it can effectively avoid the error (phase shift) from misplacement. In this work, all samples are relatively thick and strong absorption. Therefore, we mainly introduce the methods of the material parameter extraction for this type of materials.

### A. Refractive index and absorption coefficient

From the typical one-dimensional normal incidence model (air – sample - air), the received THz signal is:

$$\tilde{E}_s(\omega) = \tilde{E}_r(\omega) T_{01} T_{10} \exp(-i\tilde{n}k_0 d) \sum_{m=0}^{\infty} (R_{10}^2 e^{-2i\tilde{n}k_0 d})^m \quad (1)$$

where $\tilde{E}_r(\omega)$ is the reference signal (the optical path without sample) in frequency domain, $\omega$ is the angular frequency, $T_{01} = \frac{2}{1+\tilde{n}}$ and $T_{10} = \frac{2\tilde{n}}{1+\tilde{n}}$ are the interface transmissive coefficient (01 means from air to sample, and 10 means from sample to air), $\tilde{n} = n + i\kappa$ is the complex refractive index, $n$ is the real part of the refractive index, $\kappa = \frac{\alpha}{2k_0}$ is the extinction coefficient, $\alpha$ is the absorption coefficient, $k_0 = \frac{\omega}{c}$ is the wave number in vacuum, $d$ is the sample thickness, and $R_{10}$ is the interface reflective coefficient. For strong absorption materials without multiple reflection, the last term in Eq. (1) $\sum_{m=0}^{\infty} (R_{10}^2 e^{-2i\tilde{n}k_0 d})^m \approx 1$. The model degrades as:

$$\tilde{T}(\omega) = \frac{\tilde{E}_s}{\tilde{E}_r} \approx \frac{4\tilde{n}}{(1+\tilde{n})^2} \exp(-i\tilde{n}k_0 d) \quad (2)$$

Therefore, the refractive index and absorption coefficient can be calculated by:

$$n(\omega) = 1 + \frac{c}{2\pi\omega d}(\phi_{sam}(\omega) - \phi_{ref}(\omega)) \quad (3)$$

$$\alpha(\omega) = -\frac{2}{d}\ln\left[\tilde{T}(\omega)\frac{(n(\omega)+1)^2}{4n(\omega)}\right] \quad (4)$$

where $\phi_{sam}$ and $\phi_{ref}$ are the phase angles of the sample signal and reference signal.

**B. Dielectric response parameters**

After obtaining the complex refractive index, it is possible to calculate the dielectric constant:

$$\tilde{\varepsilon} = \varepsilon_1 + i\varepsilon_2 = \tilde{n}^2 \quad (5)$$
$$\varepsilon_1(\omega) = n^2(\omega) - \kappa^2(\omega) \quad (6)$$
$$\varepsilon_2 = 2n(\omega)\kappa(\omega) \quad (7)$$

where $\varepsilon_1$ is the real part of the dielectric constant, and $\varepsilon_2$ is the imaginary part of the dielectric constant. For describing the response of free change carriers (electrons/ions), the Drude model is introduced:

$$\tilde{\varepsilon} = \varepsilon_\infty - \frac{\omega_p^2}{\omega^2 + i\gamma\omega} \quad (8)$$

where $\varepsilon_\infty$ is the high-frequency permittivity (which is contributed from bound electrons), $\omega_p$ is the plasma frequency (which is related to free carrier density), and $\gamma$ is the scattering rate. For general industrial materials such as semiconductor and polymer composites, Drude model cannot effectively fit the experimental data. The Drude-Lorentz model should be introduced:

$$\tilde{\varepsilon} = \varepsilon_\infty - \frac{\omega_p^2}{\omega^2 + i\gamma\omega} + \sum_j \frac{f_j}{\omega_{0j}^2 - \omega^2 - i\gamma_j\omega} \tag{9}$$

where $f_j$ is the oscillator strength, $\omega_{0j}$ is the resonance frequency, and $\gamma_j$ is the damping.

## III. FIELD-MATERIAL COUPLED NEURAL NETWORK FOR MATERIAL PARAMETER EXTRACTION

Conventional deep learning techniques require amount of training datasets with labels. This is quite difficult since there are no public datasets or robustness measurement results for THz-TDS. Therefore, a field-material coupled neural network (FMCNN) is proposed in this work to address the challenge in material parameter extraction. Different from general PINN or physics-constrained neural network[39] aiming to simulate the physical field, the FMCNN here is only treated as an inverse problem solver. The schematic image is shown in Fig. 1.

In THz band, the optical and electrical properties for most materials are frequency-dependent. Therefore, the time-domain Maxwell equations are not suitable to describe the interaction between THz wave and the matter. The frequency-domain Maxwell equations can be given as:

$$\begin{cases} \nabla \cdot \tilde{D}(r,\omega) = 0 \\ \nabla \cdot \tilde{B}(r,\omega) = 0 \\ \nabla \times \tilde{E}(r,\omega) = i\omega\tilde{B}(r,\omega) \\ \nabla \times \tilde{H}(r,\omega) = -i\omega\tilde{D}(r,\omega) \end{cases} \tag{10}$$

where $\tilde{D} = \varepsilon_0 \tilde{\varepsilon}(\omega)\tilde{E}$ is the electric displacement field, $\varepsilon_0$ is the vacuum permittivity, $\tilde{E}$ is the electric field, $\tilde{B} = \mu_0 \tilde{H}$ is the magnetic flux density, $\mu_0$ is the vacuum permeability, and $\tilde{H}$ is the magnetic field. For a non-magnetic, isotropic material, the Maxwell equations can be simplified to a one-dimensional Helmholtz equation:

$$\frac{d^2\tilde{E}(z,\omega)}{dz^2} + k_0^2 \tilde{\varepsilon}(\omega)\tilde{E}(z,\omega) = 0 \tag{11}$$

Here, the partial differential equation (PDE) loss can be given as:

$$\mathcal{L}_{PDE} = \frac{1}{N_{PDE}N_\omega} \sum_{i=1}^{N_{PDE}} \sum_{j=1}^{N_\omega} \left(\frac{d^2\tilde{E}(z_i,\omega_j)}{dz^2} + k_0^2 \tilde{\varepsilon}(\omega_j)\tilde{E}(z_i,\omega_j)\right)^2 \tag{12}$$

For the optical path with a sample in THz-TDS, the optical source can be given as:

$$\tilde{E} = \tilde{E}(z_0, \omega) \tag{13}$$

Thus, the boundary condition loss for optical source can be given as:

$$\mathcal{L}_{OS} = \frac{1}{N_\omega} \sum_{i=1}^{N_\omega} (\tilde{E}(z_0, \omega_i))^2 \tag{14}$$

where $z_0$ is the location of the optical source. Similarly, the boundary condition loss for outflow boundary is:

$$\mathcal{L}_{Out} = \frac{1}{N_\omega} \sum_{i=1}^{N_\omega} (\tilde{E}(z_1, \omega_i))^2 \quad (15)$$

where $z_1$ is the location of the THz receiver. Therefore, the loss function can be written as:

$$\mathcal{L} = \mathcal{L}_{PDE} + \mathcal{L}_{OS} + \mathcal{L}_{Out} \quad (16)$$

However, the neural network is difficult to process the computation of complex values. It is possible to divide the Eq. (16) into real part and imaginary part:

$$\begin{cases} \mathcal{L}_{real} = \text{Re}\{\mathcal{L}_{PDE} + \mathcal{L}_{OS} + \mathcal{L}_{Out}\} \\ \mathcal{L}_{imag} = \text{Imag}\{\mathcal{L}_{PDE} + \mathcal{L}_{OS} + \mathcal{L}_{Out}\} \end{cases} \quad (17)$$

Since we only have the data from optical source and outflow (transmissive) boundary, it is extremely difficult to predict the complex dielectric constant during the training. Here, another neural network, dielectric constant network, should be involved as a global material property that varies only with frequency and should remain spatially uniform. To ensure physical plausibility, the network outputs are transformed via a softplus function with positive offsets:

$$\begin{cases} \varepsilon_1(\omega) = \ln(1 + \exp(\tilde{\varepsilon}(\omega))) + 1 \\ \varepsilon_2(\omega) = \ln(1 + \exp(\tilde{\varepsilon}(\omega))) \end{cases} \quad (18)$$

This guarantees that the real part of the permittivity remains greater than 1 and the imaginary part is strictly positive, consistent with passive material behavior. The complex dielectric constant is coupled in the Maxwell electromagnetic field equations. Consequently, the gradients from the PDE residual propagate to the dielectric constant network, allowing it to learn the material's frequency-dependent response without requiring direct measurement of $\tilde{\varepsilon}(\omega)$.

## IV. EXPERIMENTAL AND TRAINING SETUP

### A. Samples

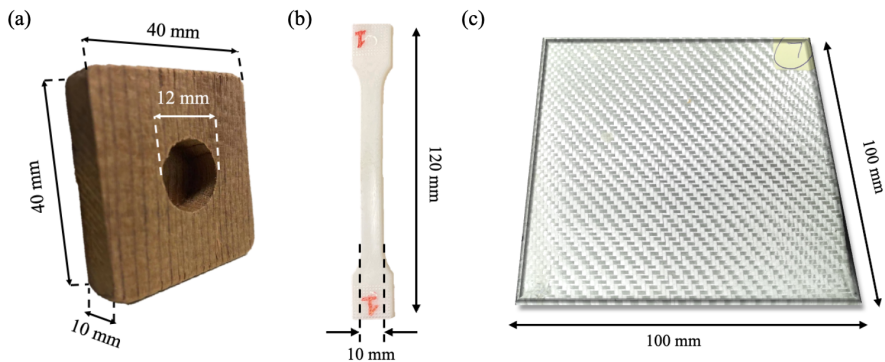

FIG. 1. (a) Spruce sample; (b) HDPE sample; (c) GFRP sample.

To validate the robustness and accuracy of the proposed FMCNN, three types of materials are selected in this work, including wood (spruce), high-density polyethylene (HDPE), and glass fiber reinforced polymer (GFRP) preform. The dimension of the spruce sample is 40 mm × 40 mm × 10 mm. There is a flat hole in the spruce sample with a depth of 7.5 mm. The diameter of this flat hole is 12 mm, as shown in Fig. 1(a). The HDPE sample is a dog-bone shape with a thickness of 6 mm, a length of 120 mm, and a center width of 10 mm. The GFRP sample was made from twill weave glass fiber fabrics featuring five 100 mm × 100 mm layers. Layers were assembled into preforms using Airtec2 spray adhesive, and were compacted under 4.6 kPa for 300 s. Yarns extended along 2 orientations in all plies namely 0° and 90°, and consolidated sample had a thickness of 1.3 ± 0.01 mm.

## B. Terahertz time-domain spectroscopy system

The schematic of the terahertz time-domain spectroscopy (THz-TDS) imaging system is shown in Fig. 2. An ultrafast femtosecond laser pulse is split into a pump beam and a probe (reference) beam. The pump beam is directed through an optical delay line to introduce a precisely controlled temporal delay and subsequently excites a photoconductive antenna (PCA) emitter to generate broadband THz pulses. The emitted THz radiation propagates to the sample, where it is reflected (or transmitted), and is then collected and focused onto a PCA detector. The probe beam, synchronized with the detector, enables coherent electro-optic sampling of the time-resolved THz electric field. The detected signal is demodulated using a lock-in amplifier to enhance the signal-to-noise ratio and recorded for further analysis.

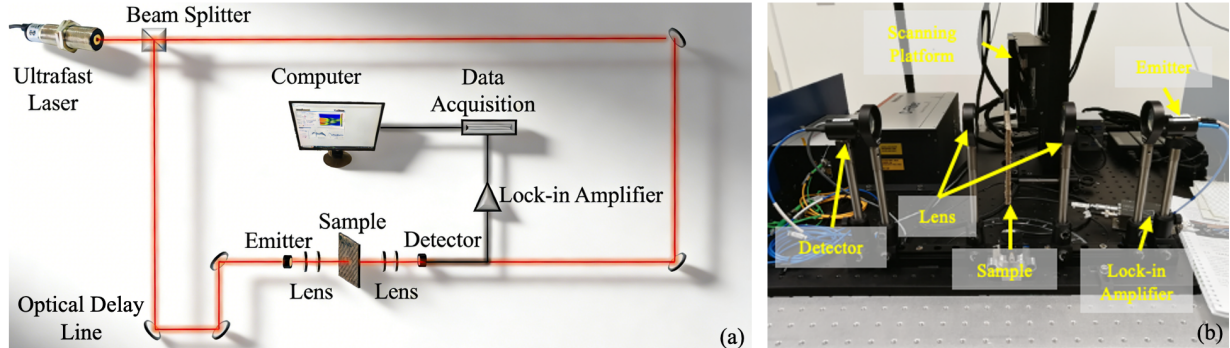

**FIG. 2.** (a) Schematic image of THz-TDS system; (b) Photograph of THz-TDS system.

In this study, a commercial THz-TDS system (Menlo Systems GmbH, Munich, Germany) is employed, providing a frequency resolution of approximately 1.2 GHz at a laser repetition rate of 100 MHz. All measurements are performed in a reflection geometry, with a spatial scanning step size of 0.5 mm.

## C. Field-material coupled neural network training

The field-material coupled neural network (FMCNN) consists of two networks: a field network and a material network. The field network comprises an input layer with 2 neurons (spatial coordinate and frequency), followed by three fully connected hidden layers, each containing 64 neurons, and an output layer with 2 neurons representing the real and imaginary parts of the electric field. The material network contains an input layer with 1 neuron (frequency), two hidden layers with 64 neurons each, and an output layer with 2 neurons representing the real and imaginary parts of the complex dielectric constant. An adaptive Tanh activation function is employed throughout both networks. Training is performed using the Adam optimizer with a learning rate of $1\times10^{-3}$ for 5000/10000 iterations, with gradients clipped to a maximum norm of 1.0 to prevent gradient explosion. The total loss function consists of two terms: a boundary-data loss enforcing the electric field at the sample interfaces (optical source and outflow boundary), and a PDE residual loss enforcing the dimensionless Helmholtz equation across randomly sampled collocation points in the spatial domain. The material network is trained through PDE residual, as the permittivity only appears in the Helmholtz equation. During training, the collocation points are randomly sampled for each iteration, while the frequency set is kept fixed. The spatial coordinates are normalized to the sample thickness, whereas the frequencies are scaled to THz units. The FMCNN framework is implemented in PyTorch and trained on NVIDIA 4060 Titan GPUs. The predicted frequency-dependent permittivity is extracted after convergence and saved for further analysis. A schematic of the training and prediction workflow is illustrated in Fig. 3.

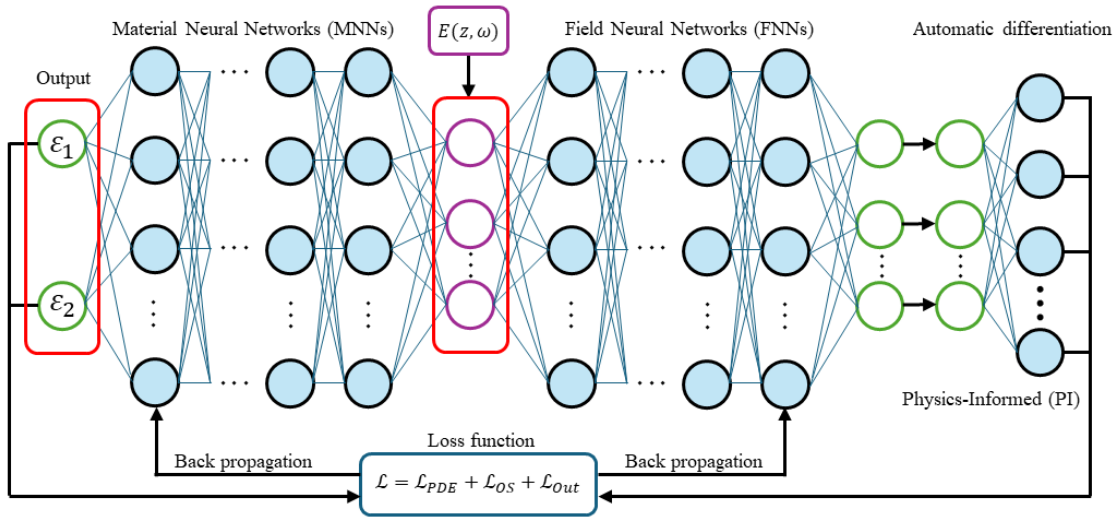

**FIG. 3.** Schematic image of FMCNN.

## V. RESULTS AND DISCUSSION

The experimental results are shown in Fig. 4. The refractive index and absorption coefficient were determined using Eqs. (3) and (4), and the complex dielectric constant was subsequently calculated according to Eqs. (5)-(7). The Drude-Lorentz model (Eq. (9)) was used to fit the calculated complex dielectric constant. Three samples with distinct material properties were tested. For the spruce sample in Fig. 4(a) and (d), the real part of the permittivity exhibits a rapid decrease around 1 THz. This behavior arises from Debye-type relaxation associated with water molecules and cellulose, which typically occurs in the 0.1-1 THz range. At frequencies above the relaxation, the molecular dipoles cannot follow the rapidly oscillating field, resulting in a reduction of the real permittivity. Consequently, the Drude-Lorentz model provides a relatively poorer fit for the spruce sample. The imaginary part of the permittivity is dominated by water relaxation absorption, which has a broad linewidth, leading to minimal variation above 1 THz. For the HDPE sample in Fig. 4(b) and (e), both the real and imaginary parts of the permittivity remain nearly constant below 2.4

THz. This behavior reflects the non-polar backbone of HDPE, which exhibits few molecular vibrations and negligible water absorption. The electronic polarization, corresponding to the high-frequency response, is nearly constant, allowing the molecules to fully follow the applied electric field. With no active dipoles or strong vibrational modes, dielectric losses are primarily due to scattering or impurities, which are negligible. The permittivity of GFRP sample is shown in Fig. 4(c) and (f). For the GFRP sample, the material is a composite of glass fibers embedded in a polymer matrix. The glass fibers are essentially dielectric, nonpolar, and exhibit very low loss in the THz range, whereas the polymer matrix (epoxy) contains polar group and shows low-THz relaxation behavior. As a result, the effective permittivity can be approximated as a weighted average, $\varepsilon_{\text{GFRP}} \approx \varepsilon_{\text{matrix}}(1 - \phi) + \varepsilon_{\text{fiber}}\phi$, where $\phi$ is the fiber volume fraction. Below 1.2 THz, the polymer chains in the matrix exhibit relaxation frequencies typically above 1 THz, and the rigid, non-polar glass fibers contribute negligible response, leading to nearly constant permittivity. At higher frequencies, vibrational resonances of the polar groups in the polymer are activated, resulting in a gradual decrease of the permittivity.

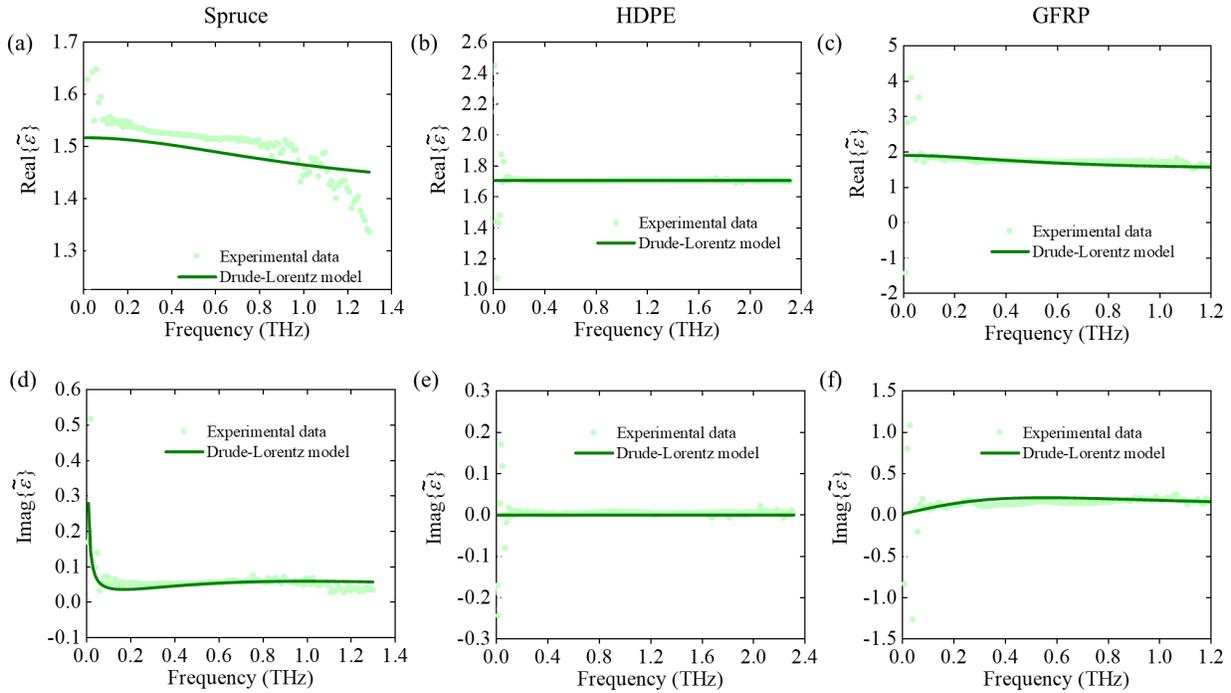

**FIG. 4.** Complex dielectric constant calculated based on one-dimensional normal incidence model (Eq. (3)-(7)) and Drude-Lorentz fitting (Eq. (9)): (a)-(c) are results of the real part from spruce, HDPE, and GFRP samples; (d)-(f) are results of the imaginary part from spruce, HDPE, and GFRP samples.

**TABLE I.** Training batch of FMCNN.

| Training batch | Sampling point number $N_f$ | Iteration/Epoch $It$ |
| --- | --- | --- |
| 1 | 0 | 5000 |
| 2 | 0 | 10000 |
| 3 | 1 | 5000 |
| 4 | 2 | 5000 |
| 5 | 5 | 5000 |
| 6 | 10 | 5000 |
| 7 | 100 | 5000 |

The proposed field-material coupled neural network (FMCNN) was employed to predict the complex dielectric constant, as shown in Fig. 5. The influence of the sampling point number, $N_f$, on the prediction results is of particular interest. The details of the training batches, including sampling point numbers and the corresponding iteration/epoch number, $It$, are summarized in Table I. Seven training configurations were considered: $N_f = 0$ with $It = 5000$, $N_f = 0$ with $It = 10000$, $N_f = 1$ with $It = 5000$, $N_f = 2$ with $It = 5000$, $N_f = 5$ with $It = 5000$, $N_f = 10$ with $It = 5000$, and $N_f = 100$ with $It = 5000$.

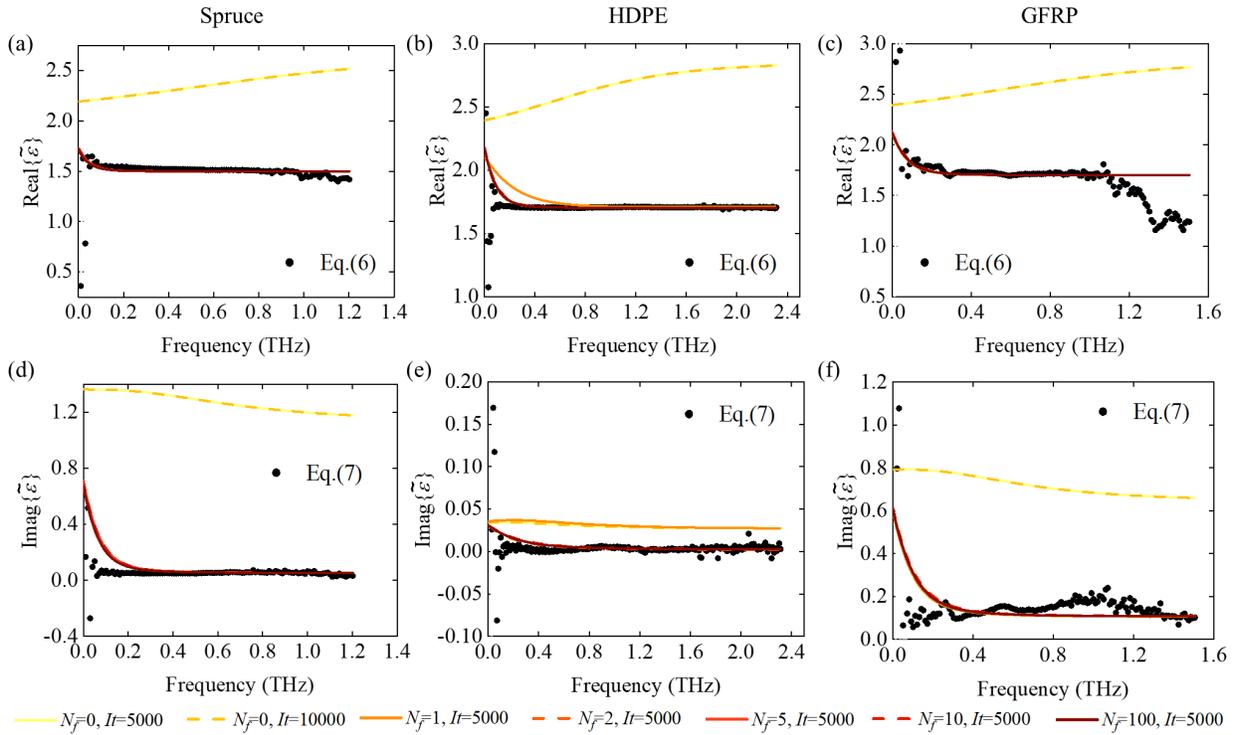

**FIG. 5.** Prediction results based on the FMCNN: (a)-(c) are results of the real part from spruce, HDPE, and GFRP samples; (d)-(f) are results of the imaginary part from spruce, HDPE, and GFRP samples.

It should be noted that when the sampling point number $N_f$ is zero, the proposed FMCNN reduces to a purely physics-constrained neural network[39], in which the PDE loss alone does not provide information for material parameter optimization. As shown in Fig. 5, in this case, both the real and imaginary parts of the permittivity exhibit large deviations from the values obtained using the one-dimensional normal-incidence model in Eq. (3)-(7). To exclude the influence of iteration number, the training was performed with $It = 5000$ and $It = 10000$, yielding nearly identical results, indicating that pure physics-constrained learning cannot predict the material parameters. Introducing even a single sampling point significantly improves the prediction trend, and increasing the number of sampling points allows the predicted complex dielectric constant to converge toward the optimum values. For instance, in Fig. 5(b) and (e), using only 1 or 2 sampling points results in relatively larger deviations of the real permittivity below 0.6 THz. These observations demonstrate that incorporating physics-informed learning effectively guides the neural network to capture the interaction between the THz wave and the material. However, further increasing the sampling point number has limited benefit, since the THz-TDS system only provides measurements at the sample boundaries, without direct observations inside the material. Therefore, in practical training, it is preferable to provide a plausive initial value for the real part of the permittivity; otherwise, the FMCNN may prioritize minimizing the PDE and data losses rather than the material parameters.

Finally, the FMCNN predictions were compared with the fitting results obtained from the Drude-Lorentz model, as shown in Fig. 6. For the spruce sample in Fig. 6(a) and (d), the real part of the permittivity shows comparable agreement between two methods, whereas the imaginary part exhibits larger deviations below 0.2 THz. This discrepancy is not due to limitations of the FMCNN, but rather arises from the finite time window, limited dynamic range, and baseline drift of the THz-TDS system, which provides reliable measurements primarily above 0.3 THz. For the GFRP sample in Fig. 6(c) and (f), the FMCNN exhibits superior performance compared to the Drude-Lorentz model above 0.2 THz. It should be emphasized that the FMCNN is not a fitting method; it is based on Maxwell's electromagnetic theory combined with neural network optimization. In contrast, the Drude-Lorentz model requires prior calculation of the complex dielectric constant and is primarily used to extrapolate material response in frequency regions that are inaccessible to the TH-TDS system.

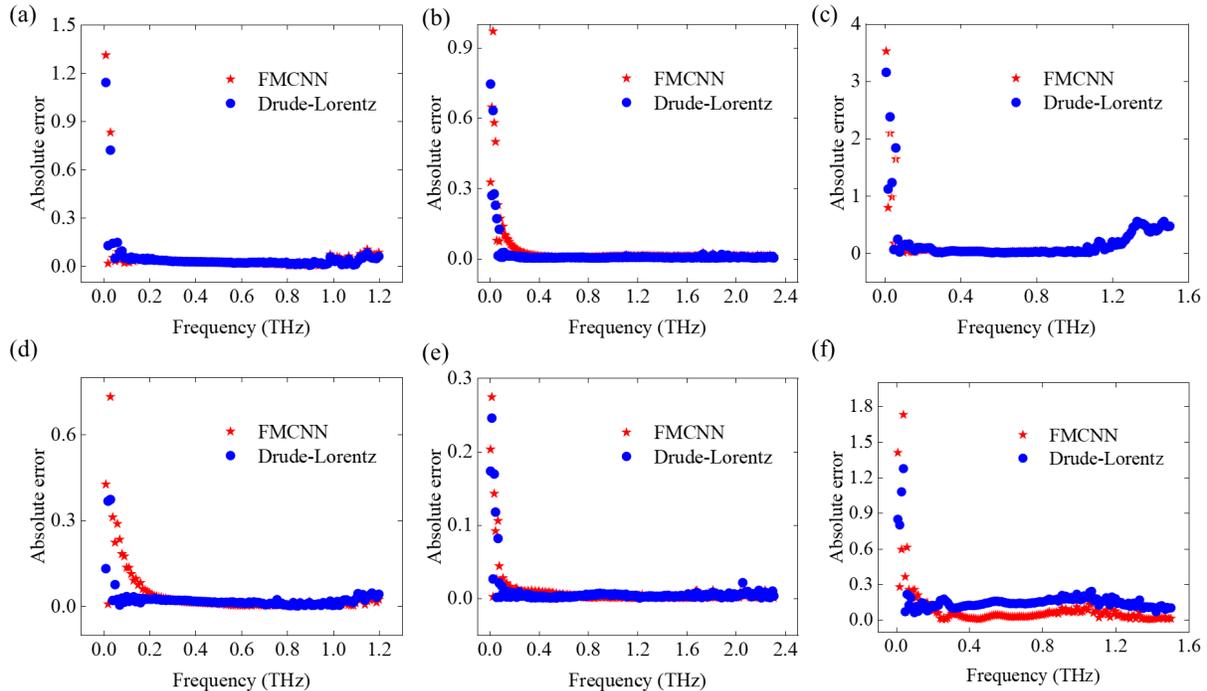

**FIG. 6.** Absolute errors between FMCNN and Drude-Lorentz model: (a)-(c) are results of the real part from spruce, HDPE, and GFRP samples; (d)-(f) are results of the imaginary part from spruce, HDPE, and GFRP samples.

## VI. CONCLUSION

In this work, a field-material coupled neural network (FMCNN) is proposed for extracting the complex dielectric constant in the THz band. The FMCNN consists of two strongly coupled components: field neural networks, which enforces the frequency-domain Maxwell equations in the form of the Helmholtz equation with PDE and boundary constraints, and a material neural network, which is embedded directly within the Helmholtz equation during training. This architecture enables the FMCNN to function as a prior-free and data-free inverse solver, requiring only measured test data as input. The method was validated by comparing the predicted complex dielectric constants with results from one-dimensional normal-incidence model and the Drude-Lorentz model. The FMCNN demonstrates high accuracy, particularly above 0.2 THz. The influence of the material neural network was further examined by varying the sampling point number. When no sampling points are provided, the FMCNN reduces to a purely physics-constrained network, which cannot reliably recover material parameters even with extended iterations. Introducing sampling points allows rapid convergence and accurate prediction of both the real and imaginary parts of the permittivity. Absolute errors indicate that the FMCNN predictions are in good agreement with Drude-Lorentz fitting, and for GFRP samples, the FMCNN achieves superior accuracy. Due to the absence of real observation points within the material, increasing the sampling point number beyond a certain threshold has limited benefit. Furthermore, material parameter extraction in the THz band is an ill-posed inverse problem, and providing a plausible initial value for the real part of the permittivity is advantageous to guide the optimization. Unlike the Drude-Lorentz model, the FMCNN does not require pre-calculated complex dielectric constants, relying instead on Maxwell's electromagnetic theory combined with neural network optimization. Overall, the proposed FMCNN provides a physics-consistent and data-efficient approach for predicting material parameters in the THz band.

## ACKNOWLEDGEMENTS

This work was supported by the Adolf Marten Fellowship (Grant n. BAM-AMF-2025-1), and Natural Sciences and Engineering Research Council (NSERC) of Canada through the Discovery and CREATE 'oN DuTy!' program (496439-2017), the Canada Research Chair in Multipolar Infrared Vision (MiViM).

## AUTHOR DECLARATIONS

### Conflict of Interest

The authors have no conflicts to disclose.

### Author contributions

**P. Zhu:** Conceptualization; Data curation; Formal analysis; Investigation; Methodology; Software; Validation; Visualization; Writing – original draft; Writing – review & editing; Funding acquisition. **X. Maldague:** Supervision; Funding acquisition; Project administration.

## DATA AVAILABILITY

The data that support the findings of this study are available from the corresponding author upon reasonable request.